\documentclass[aps,amssymb,ams,twocolumn]{revtex4}
\usepackage[latin1]{inputenc}
\pdfoutput=1
\usepackage{amsmath}
\usepackage{amsfonts}
\usepackage{amssymb}
\usepackage{graphicx}
\usepackage{color}
\usepackage{natbib}

\begin{document}

\title{Mesoscale harmonic analysis of homogenous dislocation nucleation}
\author{A. Hasan$^{(1)}$, C. E.~Maloney$^{(1)}$}
\affiliation{$^{(1)}$ Dept. of Civil and Environmental Engineering, Carnegie Mellon University, Pittsburgh, PA, USA}

\begin{abstract}
We perform atomistic computer simulations to study the mechanism of homogeneous dislocation nucleation in two dimensional (2D) hexagonal crystalline films during indentation with a circular nanoindenter. 
The nucleation process is governed by the vanishing of the energy associated with a single normal mode.
This critical mode is largely confined to a single plane of adjacent atoms.
For fixed film thickness, $L$, the spatial extent, $\xi$, of the critical mode grows with indenter radius, $R$.
For fixed $R/L$, $\xi$ grows roughly as $\xi\sim L^{0.4}$.
We, furthermore, perform a mesoscale analysis to determine the lowest energy normal mode for mesoscale regions of varying radius, $r_{\text{meso}}$, centered on the critical mode's core. 
The energy, $\lambda_{\text{meso}}$, of the lowest normal mode in the meso-region decays very rapidly with $r_{\text{meso}}$ and $\lambda_{\text{meso}}\approx 0$ for $r_{\text{meso}}\gtrsim \xi$.
The lowest normal mode shows a spatial extent, $\xi_{\text{meso}}$, which has a sublinear power-law increase with $r_{\text{meso}}$ for $r_{\text{meso}}\lesssim \xi$  and saturates at $r_{\text{meso}}\approx 1.5\; \xi$. 
We demonstrate that the $\xi_{\text{meso}}/ \xi$ versus $r_{\text{meso}}/ \xi$ curve is universal, independent of film thickness or indenter radius.
The scenario that emerges is one where the analysis of small regions, $r_{\text{meso}}\lesssim \xi$, in the material \emph{can} reveal the presence of incipient instability even when the region being probed is much smaller than the spatial extent of the critical mode.
However, the mesoscale analysis gives good estimates for the energy and spatial extent of the critical mode \emph{only} for $r_{\text{meso}} \gtrsim 1.5\; \xi$. 
In this sense homogeneous dislocation nucleation should be understood as a quasi-local phenomenon. 
\end{abstract}

\maketitle
\section{Introduction}
\label{section: introduction}

It has long been understood that plasticity in single crystals is governed primarily by the motion and production of dislocations~\cite{Hull:introDislocations}.
Typical crystals contain many pre-existing defects that can act as sources for the production of new dislocations.
However, in nanoscale samples, one can obtain essentially defect free crystals.
When subjected to inhomogeneous loading, for example underneath a nanoindenter, new dislocations will be created in these perfect crystals.
The question of where in the sample and under what conditions this happens is still surprisingly contentious.  

One naive expectation is that the production of new dislocations on a particular slip system is governed by the resolved shear stress, much in the same way that the motion of a pre-existing dislocation is known to be governed by the resolved stress~\cite{Gouldstone:2001eu}.
Li and co-workers~\cite{Li:2004JMPS} first pointed out that this simple and intuitive idea is not correct and demonstrated this quantitatively in atomistic models.
Instead, they proposed to use a modified version of a stability criterion first given by Hill in the context of a continuous elastic medium~\cite{HILL:1962by}, the so-called $\Lambda$-criterion~\cite{Li:2002ib}.

Later, Miller and co-workers~\cite{Miller:2004ad, Miller:2008qw} showed that, although the $\Lambda$-criterion worked for some potentials and some indentation geometries, it failed for others.
The $\Lambda$ value could indicate instability at many locations in the material even while the system remained mechanically stable.
Furthermore, when the system did become unstable, the motion responsible for triggering the instability, the so-called dislocation embryo, was localized in a region far from those atoms considered least stable in terms of the $\Lambda$-criterion.
Miller and Acharya~\cite{Miller:2004ad} initially proposed a new stability criterion derived from Field Dislocation Mechanics (FDM) theory, but later Miller and Rodney (MR)~\cite{Miller:2008qw} showed that this criterion had its own shortcomings.

Most recently, MR~\cite{Miller:2008qw} proposed to examine the energy eigenmodes of regions of space extended over many atoms, but smaller in spatial extent than the full system.
In particular, since the dislocation embryo was found to be localized largely on a disk-like region, they chose disk-like regions of adjacent planes of atoms {\it a posterior} to have a large enough spatial extent to encompass a whole embryo.  
They found that the lowest energy eigenvalue of the disk-like region was an excellent predictor of the location of the embryo.
Since then, others have used similar approaches to study various types of nucleation phenomena~\cite{Delph:2009vf, Delph:2010zm, Delph:2011}.

This initial work left several important issues open.
On one hand, MR showed evidence for the growth of the embryo with the radius of curvature of the indenter tip.
On the other hand, they did not discuss how the choice of \emph{size} of the disk-like meso-region would affect the analysis.
One might wonder, for instance, if a meso-region would be able to detect the instability if it were much much smaller than the intrinsic embryo.
An important related question is how rapidly the lowest energy eigenvalue approaches zero with increasing meso-region size.

In this work, we show, in agreement with MR, that the intrinsic dislocation embryo grows with the radius of the indenter.
However, here, we focus on films of finite thickness and show that even when the ratio of film thickness to indenter radius is kept constant, the embryo, surprisingly, grows with the film thickness.
One can argue dimensionally that this effect would be completely absent in any continuum description, as in reference~\cite{Li:2004JMPS}, that does not contain an additional length scale in addition to the film thickness and indenter radius. 
We also show that a meso-scale analysis can detect the presence of a dislocation embryo even when the meso-region is much much smaller than the embryo itself.
In particular, we show that the inferred spatial extent of the embryo captured by the meso-region becomes precisely equal to the true spatial extent of the embryo, but only after the meso-region fully encompasses the embryo.

The rest of this paper is organized as follows. 
In section~\ref{section: sim protocol } we describe the details of our simulation setup and our athermal, quasitatic indentation protocol. 
In section~\ref{section: kinematic description}, we give a give a brief kinematic description of the mechanism of homogeneous dislocation nucleation. 
In addition, we precisely define our measure for the embryo size, $\xi$, and present our results on its scaling properties.
Section~\ref{section: meso analysis} contains our rationale and results on the mesoscale analysis.
In particular we discuss how the lowest meso-scale eignevalue and the inferred spatial extent of the dislocation embryo scale with meso-region size.  
Finally in section~\ref{section: discussion} we conclude with a brief summary and discussion on implications of our results and an outline of future work. 

\section{Simulation Protocol and Formalism}
\label{section: sim protocol }

We perform nano-indentation simulations via energy minimization dynamics.
The LAMMPS molecular dynamics framework~\cite{lammps} is used to perform the simulations. 
The system is two dimensional (2D) with a Lennard-Jones (LJ) interaction potential. 
All energies and distances are measured in Lennard-Jones units. 
We use a cutoff of $2.5$ for the interactions. 
Minimization stops when no force on any particle exceeds $10^{-8}$.

The system is periodically replicated on the sides and bounded by a half-space of immobile particles at the bottom that interact with bulk particles through identical LJ interactions. 
The nano-indenter is modeled by a circular arc of LJ particles (with spacing equal to the lattice constant) that are constrained to move as a rigid body. 
The hexagonal crystal is prepared with first nearest neighbors placed at the equilibrium bond spacing and subsequently allowed to relax. 
The initial quenched stresses are much smaller than the stresses induced by the subsequent indentation. 
The three crystal axes are placed at $\pm n\pi/3$ with respect to horizontal.

\begin{figure}[ht]
\begin{center}
\includegraphics[width=.15\textwidth]{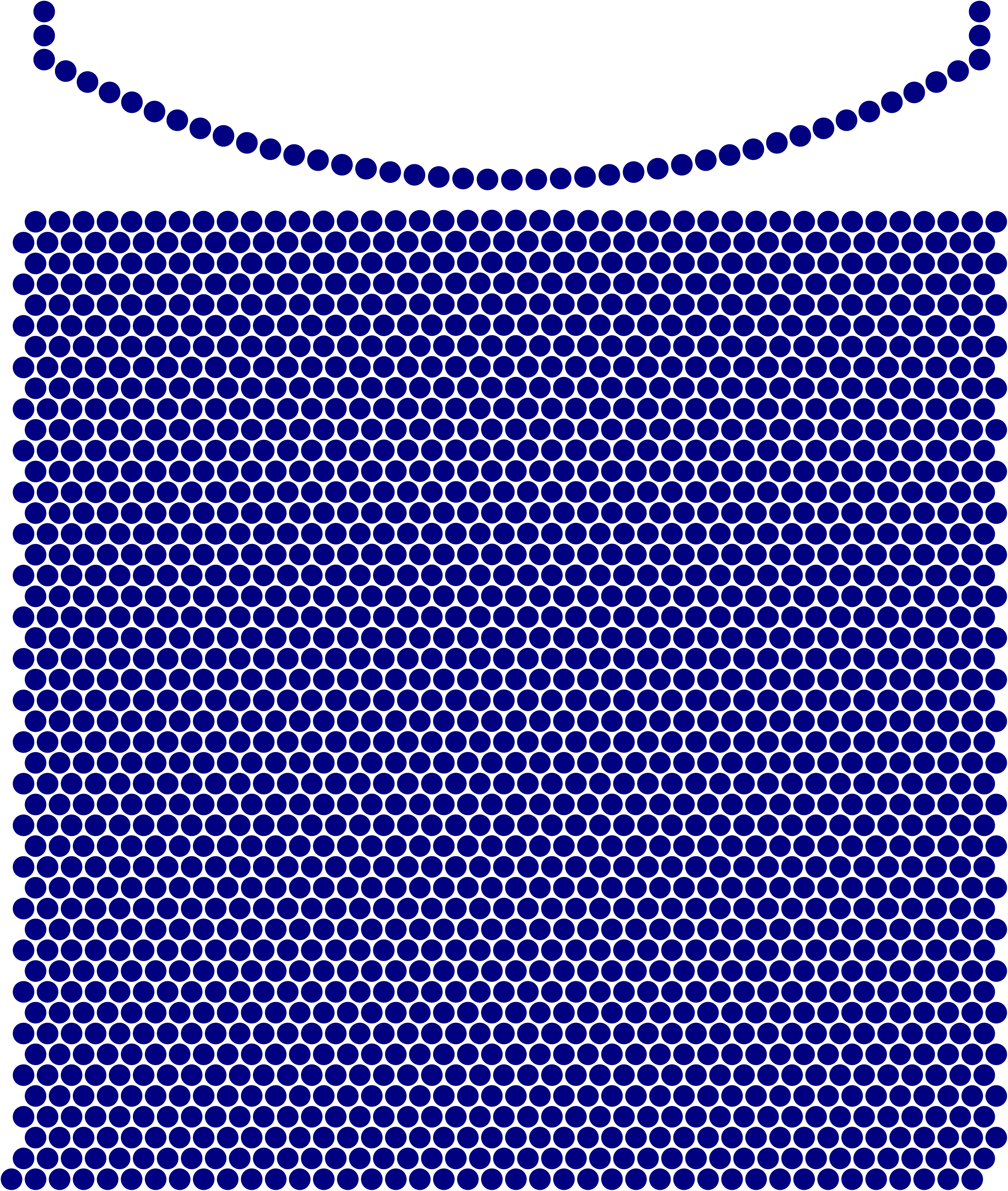}
\includegraphics[width=.15\textwidth]{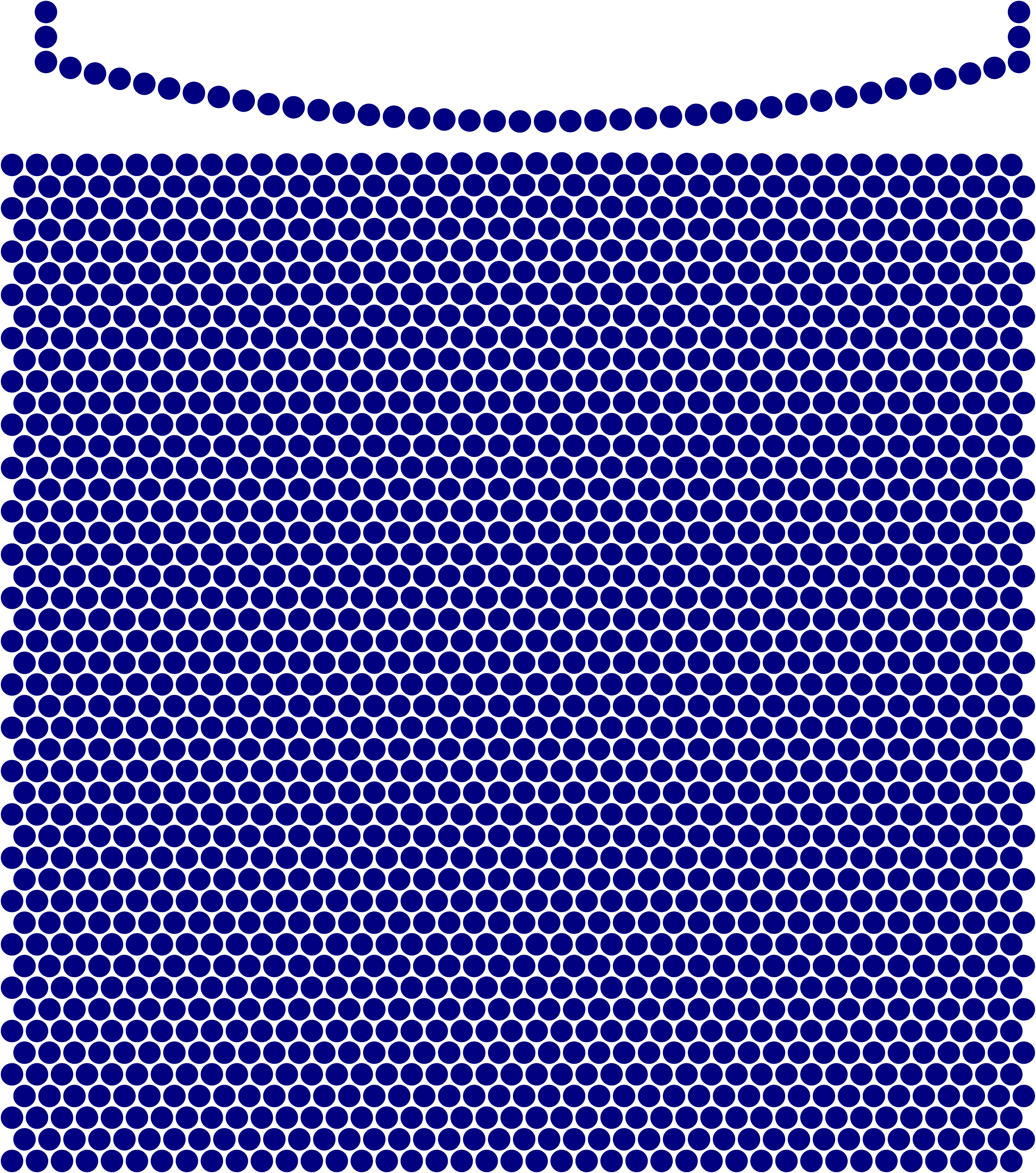}
\includegraphics[width=.15\textwidth]{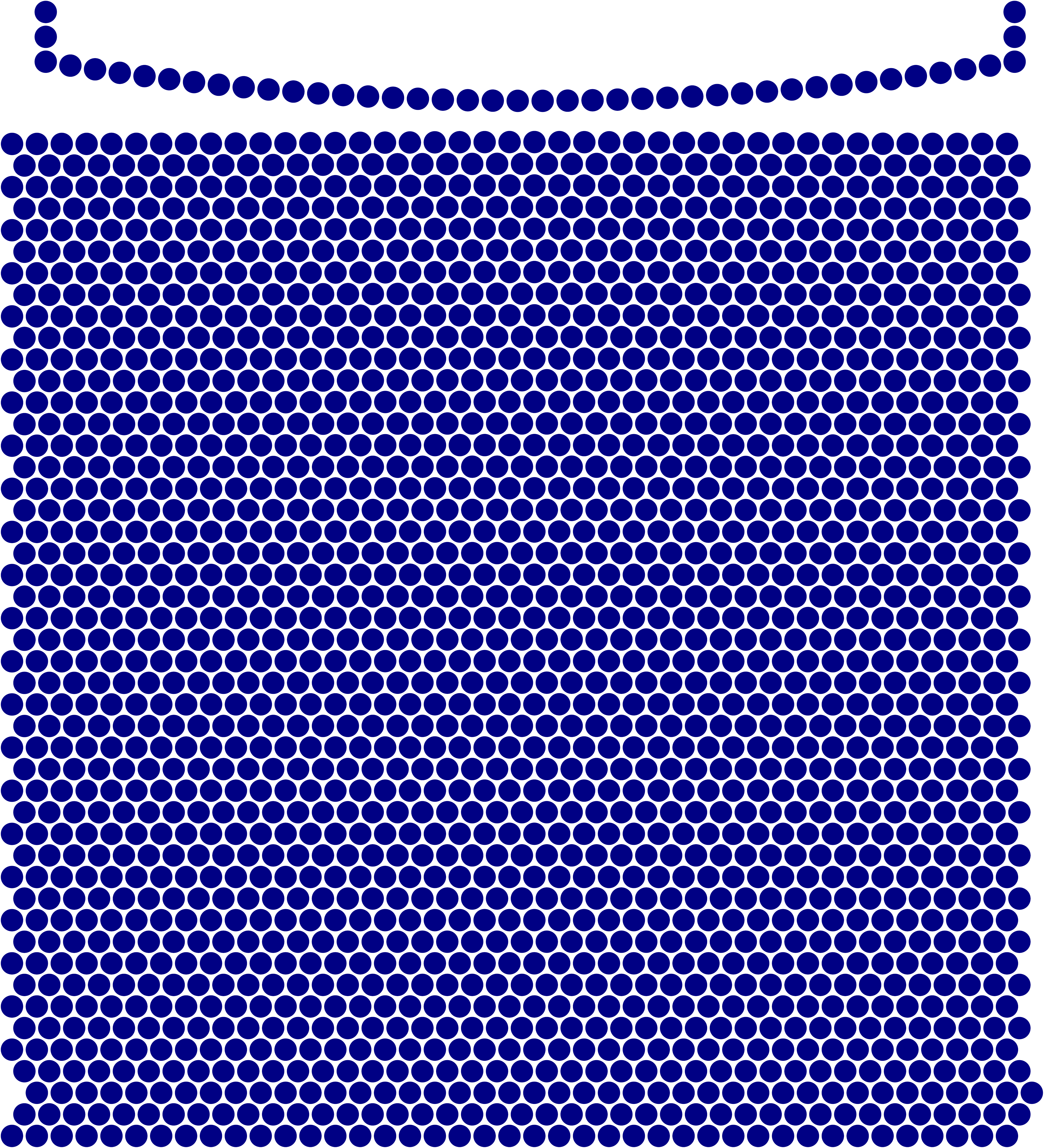} 
\includegraphics[width=.15\textwidth]{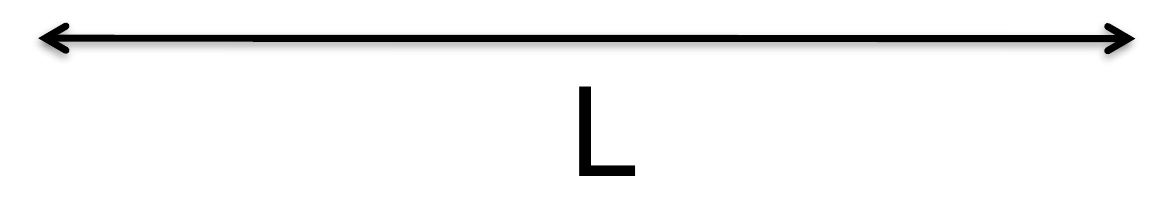}
\end{center}
\caption{ Schematic of our 3 indenter-film geometries. The width and depth in each case are equal to $L$. 
Left: $R_{\text{indenter}}=L$.
Center: $R_\text{{indenter}}=2L$.
Right: $R_{\text{indenter}}=3L$.
}
\label{fig: three geometries}
\end{figure}

\begin{figure}[ht]
\begin{center}
\includegraphics[width=0.5\textwidth]{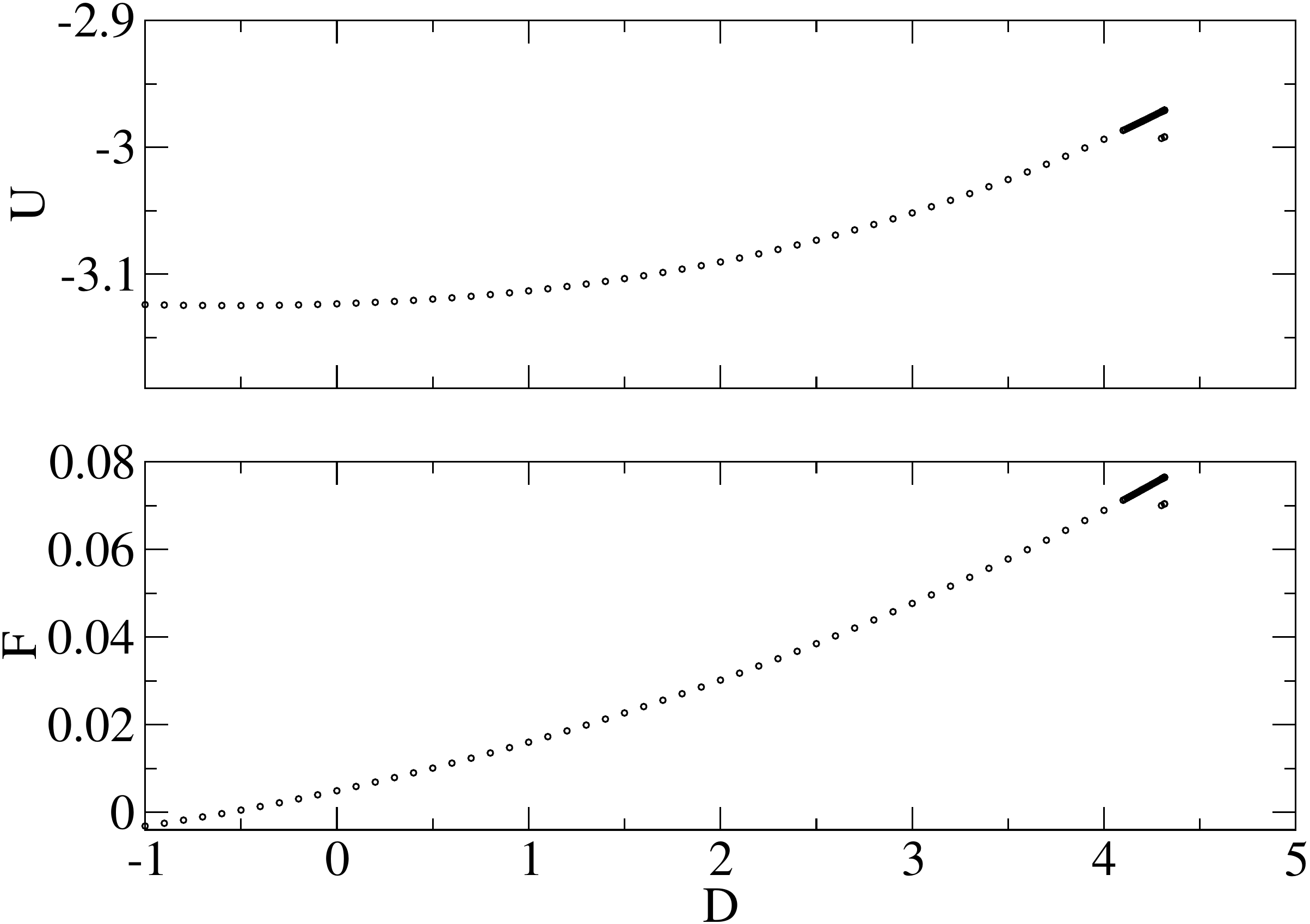}
\end{center}
\caption{ 
Top: Potential energy, $U$, of the crystalline film as function of indenter depth.  
Bottom: Load, $F$, on the indenter atoms in the vertical direction as a function of indenter depth. 
}
\label{fig: energy and load}
\end{figure}

For our 2D system there are three geometrical parameters: the crystalline film's width and thickness and the indenter radius. 
For the results in this paper, the systems are all of aspect ratio (ratio of width to thickness) 1:1. 
We verified that if we double the width while keeping the same film thickness (an aspect ratio of 2:1), our results don't change. 
We denote the film thickness by $L$. 
In figure~\ref{fig: three geometries} we sketch the three geometries studied: for each $L$, we perform simulations for 3 different indenter radii, $R$, which are equal to $L$, $2L$ and $3L$.   

Following standard procedures, we simulate the athermal, quasi-static (AQS) deformation by alternately moving the indenter by a discrete step and minimizing the energy~\cite{Maloney:2006oc}.  
When the system loses mechanical stability and a dislocation loop (or edge dipole in 2D) is nucleated, the energy and indenter load will decrease at fixed indenter depth. 
This energy decrease defines an event. 
In figure~\ref{fig: energy and load} we show the total potential energy of the system and the load on the indenter during a typical AQS simulation as a function of the indenter depth, $D$.  
At the depth, $D=0$, the indenter tip just reaches the undisturbed top surface of the crystal.
The system loads and gains energy reversibly during indentation until a drop in the energy at constant indenter depth that corresponds to an irreversible dislocation nucleation event.
To study the nucleation process in detail, we restart the simulation in a configuration just before the nucleation event and restart the stepping protocol with decreased step size to converge to the nucleation point with a precision limited only by our energy minimization algorithm. 

 The second derivative of the energy with respect to particle positions, or Hessian matrix, is defined as: $H_{i\alpha j\beta} = \frac{\partial^2 U}{\partial x_{i\alpha}\partial x_{j\beta}}$.
 We use latin to index particle number and greek to index cartesian components. 
 At any particular indenter depth we compute the few smallest energy eigenvalues of the Hessian (and corresponding  eigenvectors) using Krylov subspace based routines in the MATLAB eigs() function with default parameters.

\section{Kinematic description of dislocation nulceation}
\label{section: kinematic description}

Energy minimization dynamics cause the system to follow local minima of the atomistic potential energy surface (PES) as they move in configuration space under changes in $D$. 
The PES itself changes smoothly with $D$, however, topological changes in its stationary points occur. 
In particular, incrementing $D$ may cause flattening of the PES along some direction~\cite{Lacks:1998prl}.
In the simplest scenario, the system suffers a mechanical instability, such as a homogenous dislocation nucleation, when the PES has a single normal mode with a vanishing eigenvalue.
Generically, to lowest non-trivial order, the potential energy surface (PES) of an atomistic system near the loss of mechanical stability should be given by a cubic term in the reaction co-ordinate plus a term which is bi-linear in the reaction co-ordinate and the external control parameter, $D$~\cite{Maloney:2006Lacks}. 
Such a scenario is known as a saddle-node bifurcation or fold catastrophe in control theory. 
The bifurcation should be of saddle-node type as long as there are no symmetries in the system and there is only a single degree of freedom in the high dimensional landscape along which the curvature of the PES vanishes as the system is driven externally.

\begin{figure}[t] 
\includegraphics[width=0.5\textwidth]{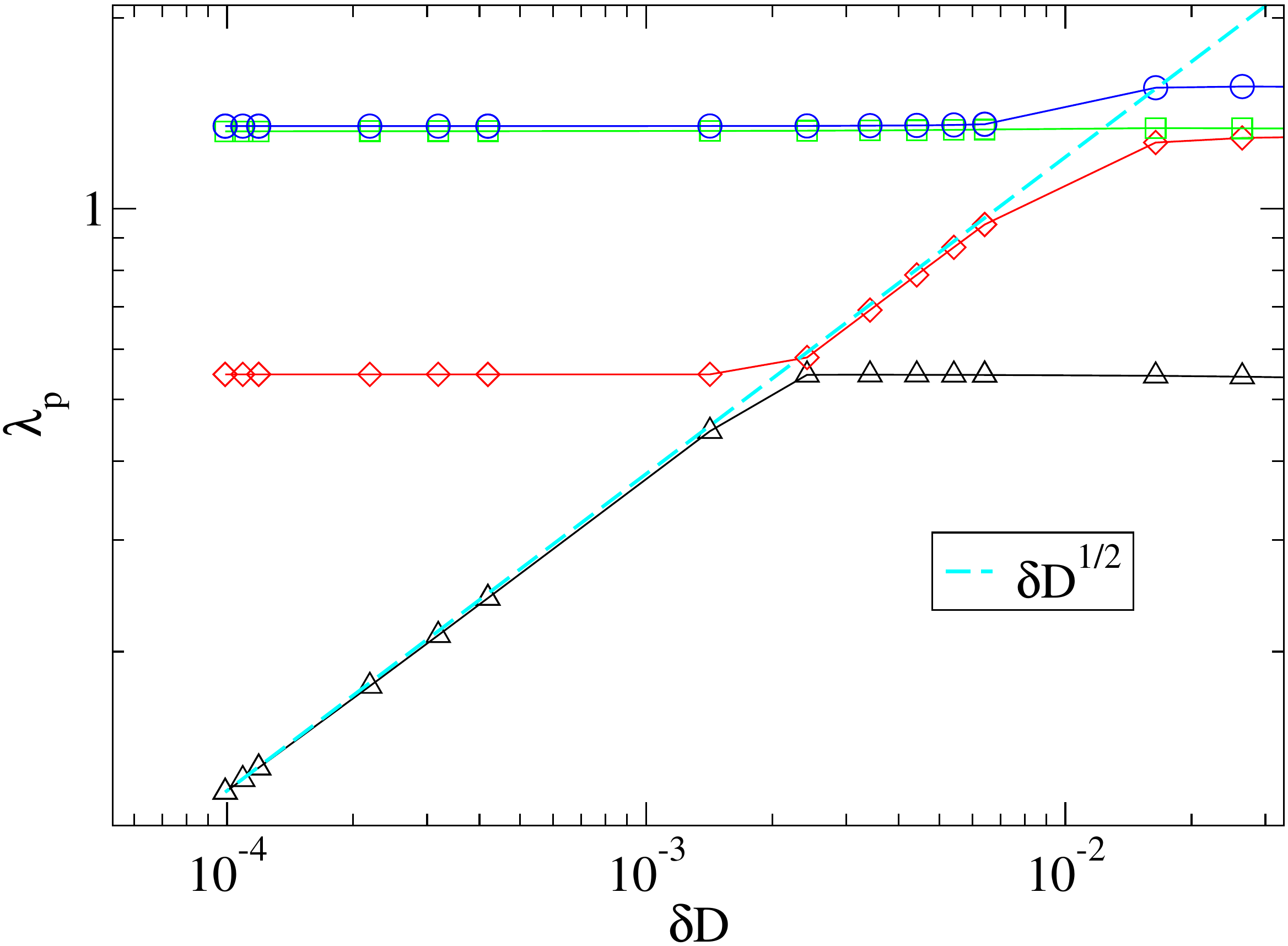}
\caption{
Smallest 4 energy eigenvalues, $\lambda_{p}$, of one of our systems ($L=45$ and $R_{\text{indenter}}=3L$), as function of $\delta{}D = D_{c} - D$. The dashed (cyan) line tracks the critical normal mode along which the system is driven to instability. 
}
\label{fig: lowest 4 eigenvalues}
\end{figure}	

\begin{figure}[th!]
\begin{center}
\includegraphics[width=.34\textwidth]{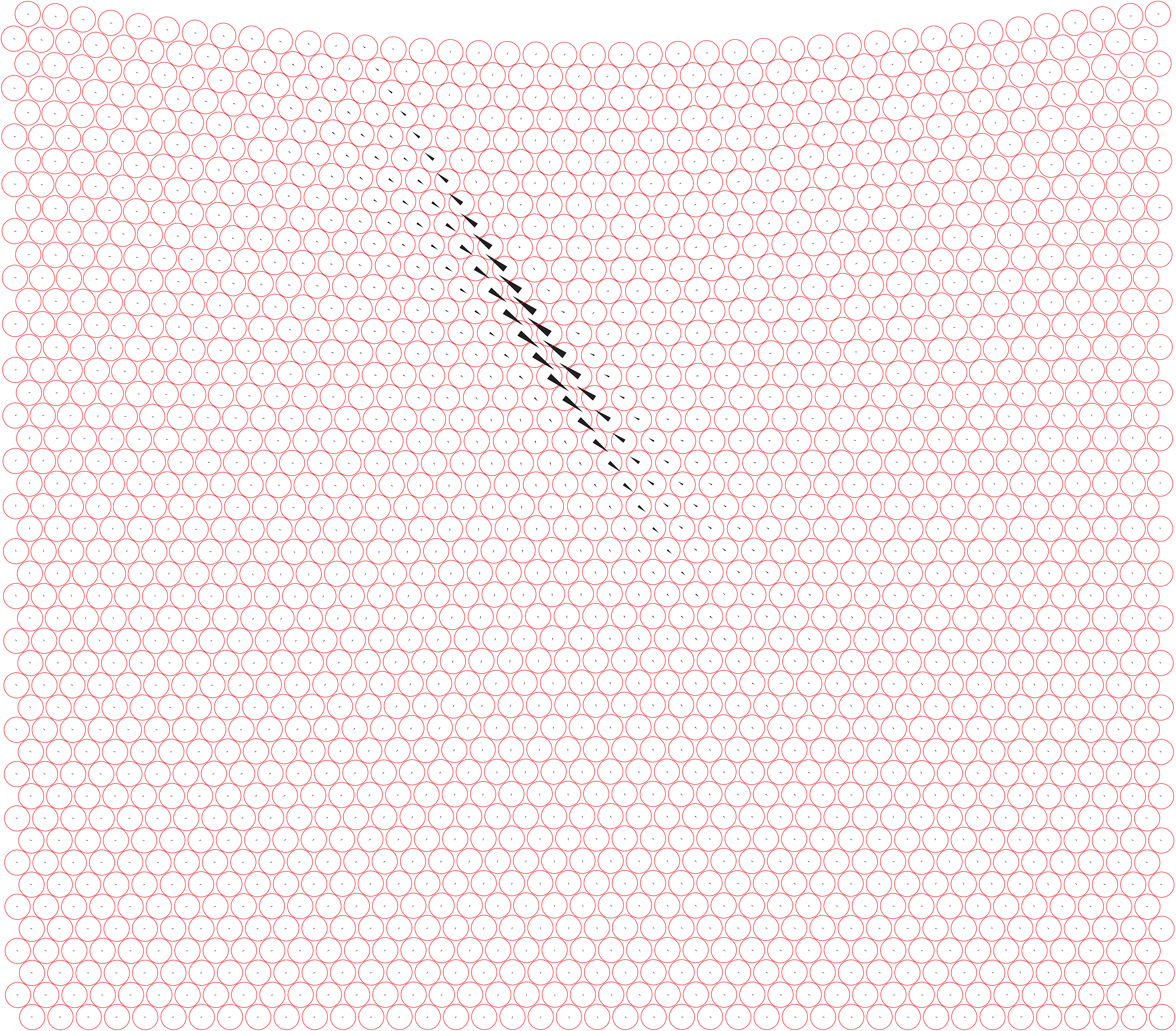}
\includegraphics[width=.31\textwidth]{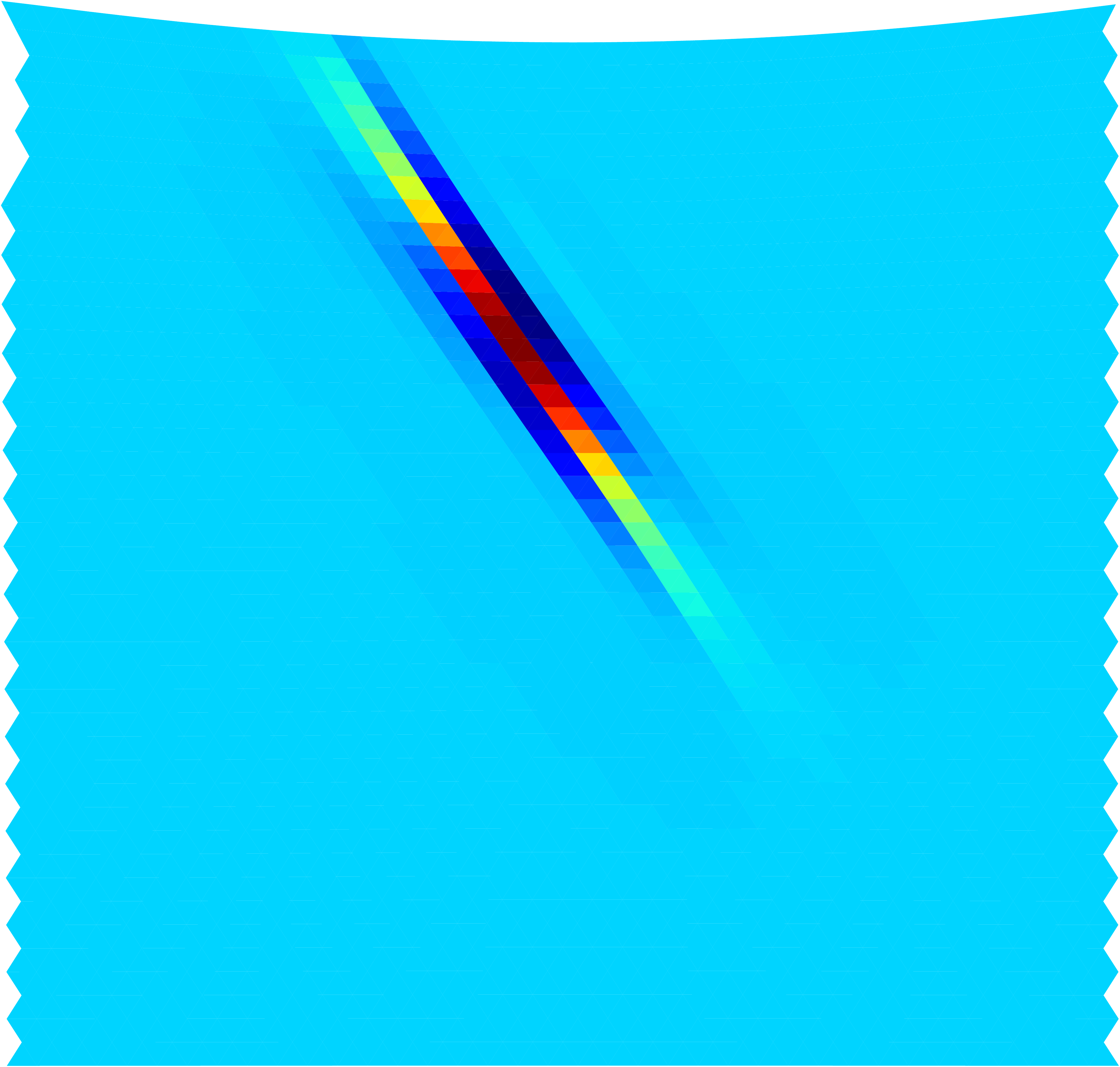} 
\includegraphics[width=.31\textwidth]{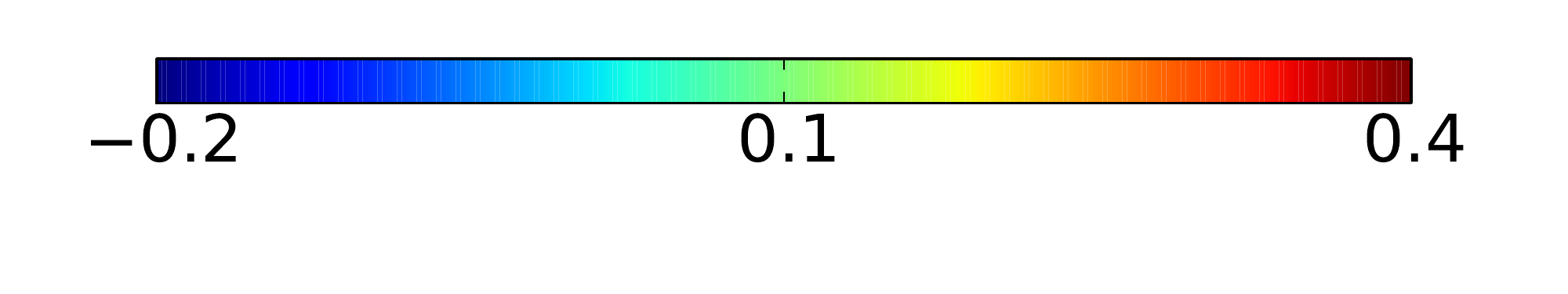} \\
\end{center}
\caption{
Top: Normal mode corresponding to lowest eigenvalue at: $\delta D=D-D_c \sim10^{-6}$, for the system $L=45$, $R=3L$. 
Center: $\Omega$ field computed from the normal mode with $\hat{s} \approx 2\pi/3$. 
}
\label{fig: omega field}
\end{figure}

In our simulations we see that the system is driven to instability along a single normal mode and the loss of mechanical stability coincides with nucleation of a single pair of dislocations~\cite{Hasan:2012bif}. 
We denote $D_{c}$, the \emph{critical depth}, as the depth of the indenter where a dislocation pair nucleates. 
In figure~\ref{fig: lowest 4 eigenvalues}, we show the lowest energy eigenvalues of one of our systems as a function of $\delta{}D$ (where $\delta{}D=D_{c} -D$). 
As expected from saddle-node bifurcation theory \emph{one} eigenvalue descends through the spectrum like $\sqrt{\delta{}D}$. 
This behavior is typical of all our systems.

We call the normal mode corresponding to the eigenvalue vanishing as $\sqrt{\delta{}D}$, the \emph{critical} mode.
At the bifurcation point, this critical mode points, in the configuration space, along the reaction coordinate along which the system is driven to instability. 
Although saddle-node bifurcation theory gives no indication of the spatial structure of the critical mode, it tells us when the lowest eigenmode in the system gives an accurate representation of the reaction co-ordinate for the pair nucleation event, {\it i.e.}, the dislocation embryo.
For the system under study in figure~\ref{fig: lowest 4 eigenvalues}, the direction on the PES that is flattening -- corresponding to the embryo -- only corresponds to the lowest eigenmode for $\delta D \leqslant 10^{-3}$ and the lowest normal mode is only a meaningful proxy for the embryo below this $\delta D$. 
In the analyses and results below, we always verify explicitly that we are close enough to the bifurcation point so that the lowest eigenvalue is descending through the spectrum according to the saddle-node bifurcation analysis.

Figure~\ref{fig: omega field} (top), shows the critical mode in a typical small system. 
Dislocation dipole nucleation is characterized by the anti-parallel motion of a small number of atoms on adjacent crystal planes.
As reported by MR, we find that the embryo involves more than the neighborhood of a single atom: it is \emph{nonlocal}.
Following the spirit of MR but departing in the particular details, we quantify the jump in the mode vector by the following procedure.
First we triangulate the lattice.  
Then, on each triangle, we make a linear interpolation of the mode vector defined at each of the three nodes.
Next we select a particular crystal axis to analyze and choose only the component of the mode vector along that axis.
Finally we take the derivative of that quantity in the direction normal to the crystal axis of interest.
This quantity is called $\Omega$.
 
Figure~\ref{fig: omega field}(top) shows a typical critical mode, and figure~\ref{fig: omega field}(bottom), shows the associated $\Omega$ field corresponding to the $\theta=-\pi/3$ crystal axis.
As expected from figure~\ref{fig: omega field}(top), $\Omega$ is practically zero everywhere except in the core of the embryo. 
The embryo center is defined as the centroid of the triangle with the greatest value of $\Omega$. 
We define the slip-line as the line along the axis of interest containing the embryo center.
Given the extended region of non-zero $\Omega$ along the slip-line, it does not make sense to talk of any \emph{point} of nucleation. 
Rather, the nucleation event is spatially extended.

\begin{figure}[t!]
\begin{center}
\includegraphics[width=.5\textwidth]{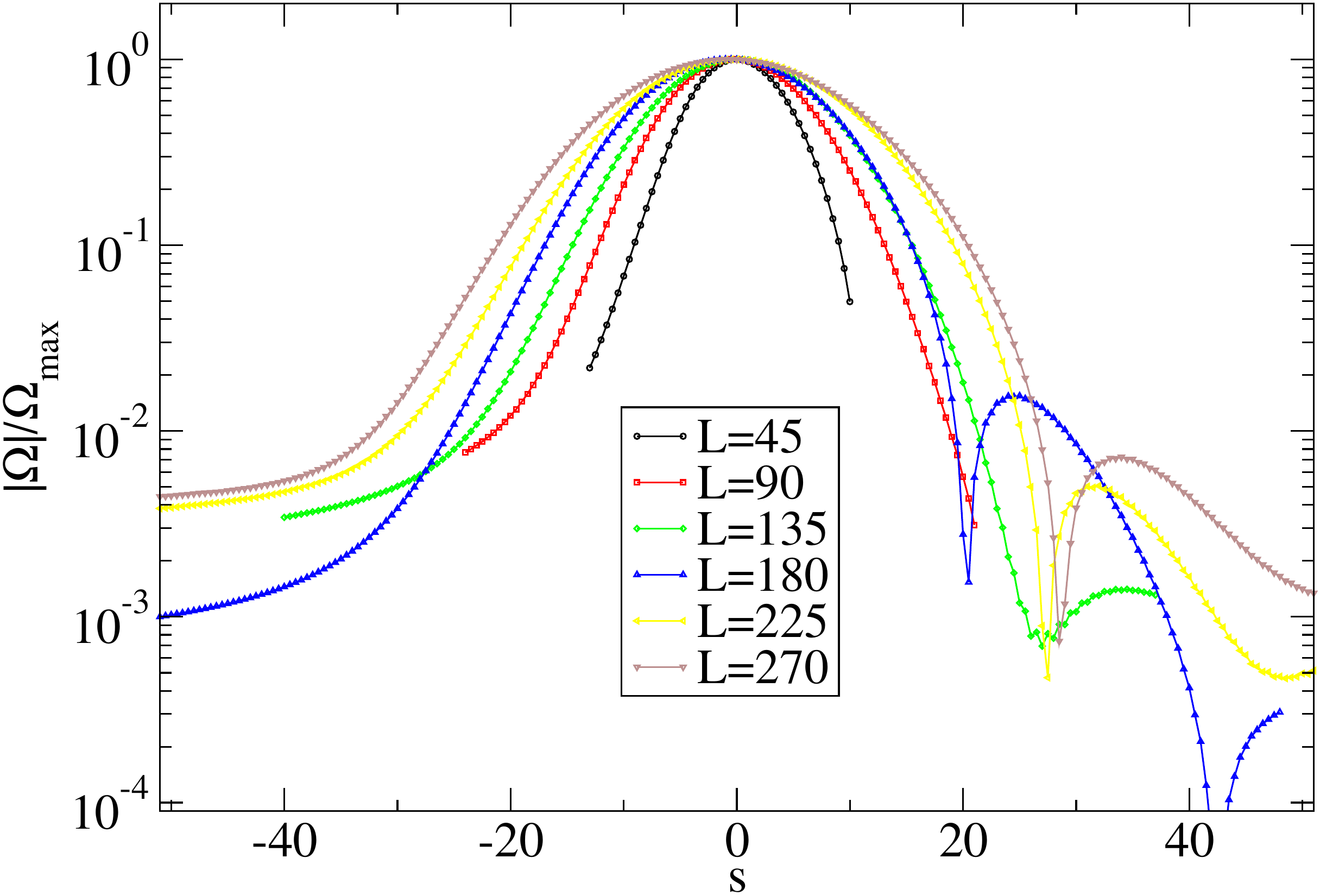}  \\
\vspace{0.01\textwidth}
\includegraphics[width=.5\textwidth]{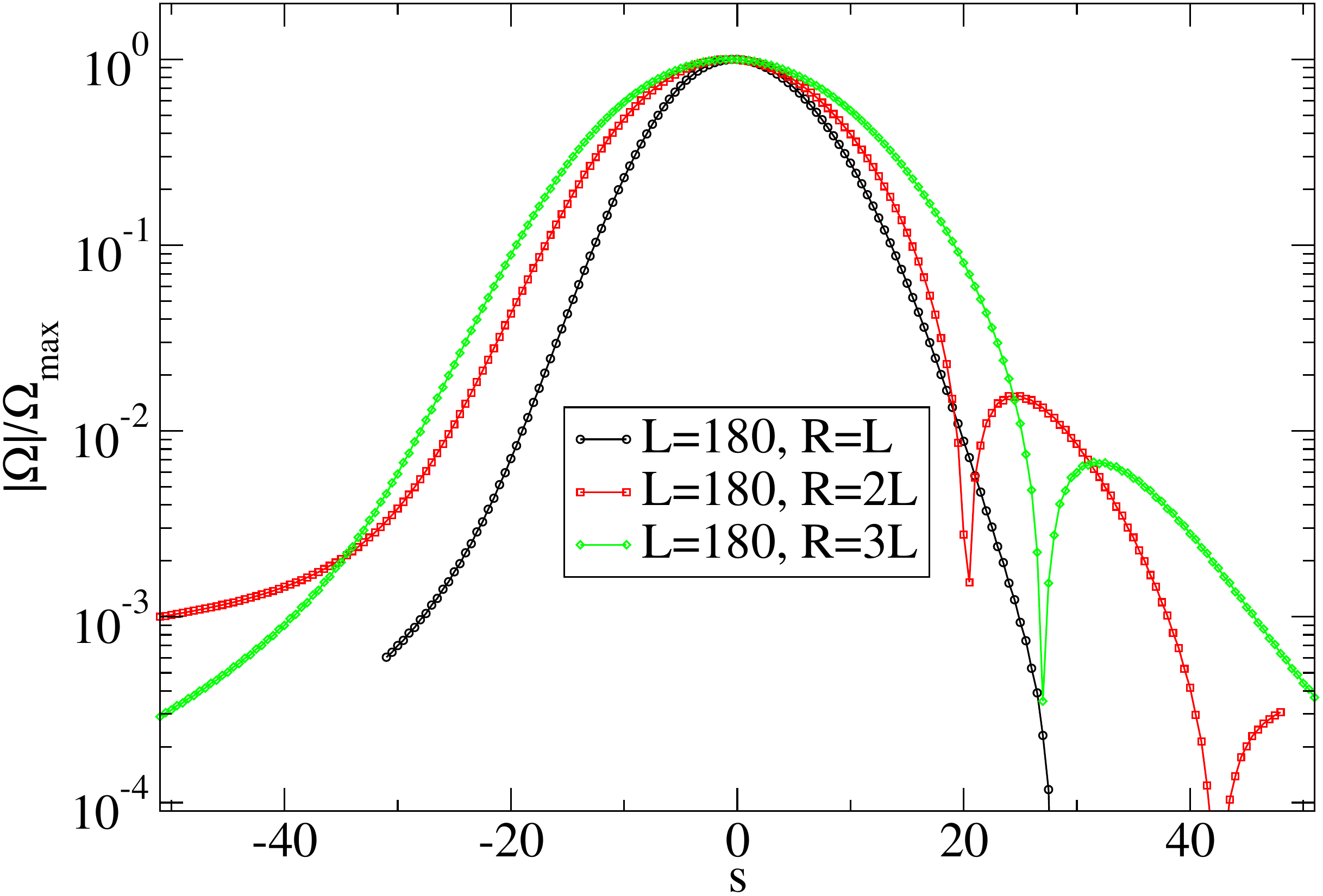}  \\
\end{center}
\caption{ $\Omega(s)$ curves on slip-planes, scaled by their maximum value $\Omega_{\text{max}}$. 
Top: For different different film thicknesses $L$ and one geometry, $R_{\text{indenter}}=2L$.
Bottom: For $L=180$ and $R_{\text{indenter}}=L,2L,3L$. }
\label{fig:omega curves}
\end{figure}

\begin{figure}[t]
\includegraphics[width=0.5\textwidth]{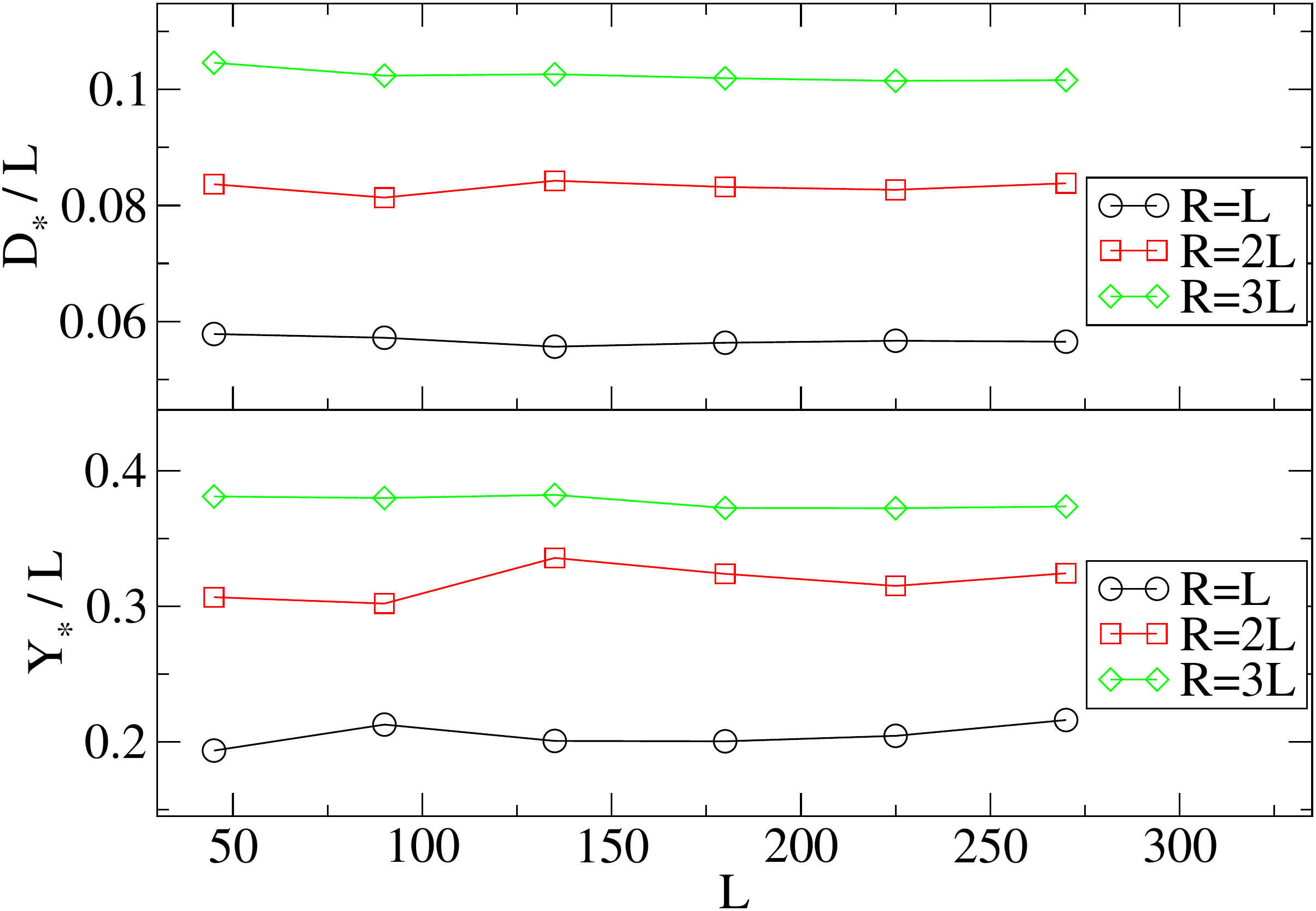}
\caption{
Top: Depth of indenter  in the last stable configuration, $D_{*}$, scaled by film thickness, $L$, for all geometries..
Bottom: Depth of the defect center, $Y_{*}$, scaled by film thickness, $L$, for all geometries.
(All depths are measured from the undeformed top surface of the crystal.)
}
\label{fig: nucleation depth}
\end{figure} 

In figure~\ref{fig:omega curves}, we extract and plot $\Omega(s)$ profiles along the slip-line, for systems of thickness L = 45, 90, 135, 180, 225, 270 and for all 3 indenter-crystal geometries (shown in figure~\ref{fig: three geometries}).
$s$ indicates distance along the slip-line with $s=0$ indicating the point of maximum $\Omega$.
$s > 0$ corresponds to locations closer to the surface and $s<0$ corresponds to locations closer to the substrate. 
The overall normalization of the critical eigenmode is arbitrary, and the $\Omega$ curves are all normalized by their maximum value.
Figure~\ref{fig:omega curves} (top) gives $\Omega(s)$ for various overall film thickness, $L$, but for a fixed indenter-crystal geometry ($R=2L$).
Figure~\ref{fig:omega curves} (bottom) gives $\Omega(s)$ for various indenter radius at a fixed film thickness of $L=180$. 

The $\Omega(s)$ profiles of all systems show smooth decay from their maximum. 
For all $L$ and $R$ they are well approximated by a Gaussian profile near the center. 
To define an embryo size, we fit the $\Omega(s)$ profiles, from their peak to half their maximum value,  to Gaussian functions of the form: 
\begin{equation} \label{eq:defintion of xi}
\Omega(s)/\Omega_{\text{max}} = e^{-s^{2}/\xi^2} 
\end{equation}
It is apparent from the $\Omega(s)$ profiles of figure~\ref{fig:omega curves} that $\xi$ increases with $L$ for fixed $R/L$ and $\xi$ increases with $R$ at fixed $L$. 

Before looking at the variation of $\xi$ with $L$, it is instructive to look at the scalings of a couple of more elementary quantities.
We denote the depth of the defect-center (the centroid of the nearest neighbor triad with the highest $\Omega$) measured from the top surface of the undisturbed crystal as $Y_{*}$.
The indenter depth of the last stable configuration measured from the same reference as before is denoted by $D_{*}$. 
Figure~\ref{fig: nucleation depth} shows $D_{*}$ and $Y_{*}$ scaled by film thickness, $L$, for all three indenter-crystal geometries.
To a very good approximation these quantities scale linearly with $L$.
This simple linear scaling seems to suggest that a length-scale free continuum model might suffice to describe dislocation nucleation. 
However we show below that $\xi$ scales in a much more complicated way with $L$ than $D_*$ and $Y_*$.  

\begin{figure}
\begin{center}
\includegraphics[width=0.5\textwidth]{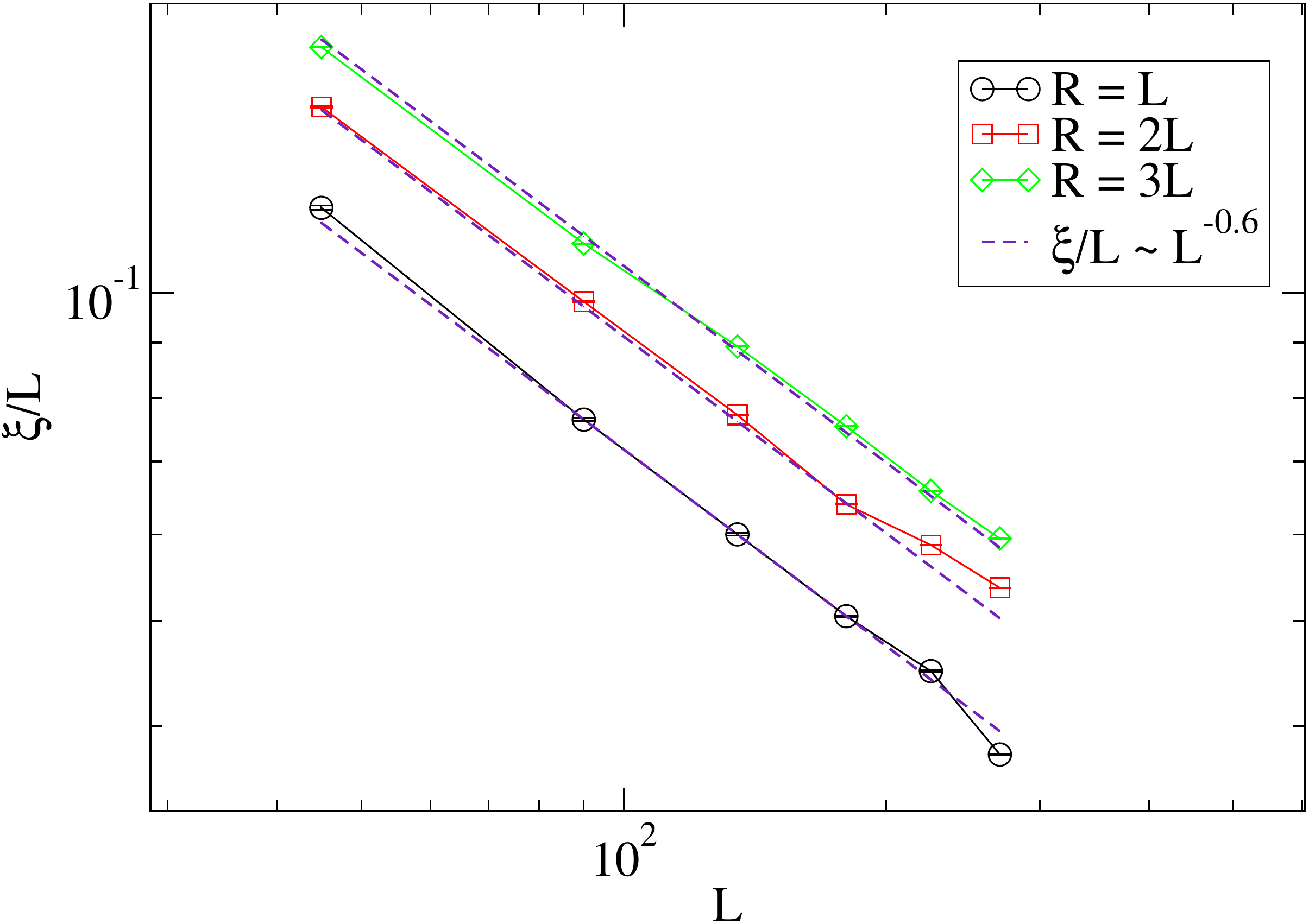}
\end{center}
\caption{ 
$\xi/L$ as function of film thickness, $L$, for all 3 indenter-crystal geometries obtained from $\Omega(s)$ curves.
}
\label{fig: xi scaling}
\end{figure}

In figure~\ref{fig: xi scaling}, we present the value of $\xi$, scaled by film thickness, $L$, for each of our 3 indenter-crystal geometries. 
We estimated $\xi$ using the generalized linear least squares method~\cite{Rice:StatsBook}, to log $\Omega(s)$ curves upto half their maximum value.  
We have checked that our results were not sensitive to the details of the fitting procedure.  
To a good approximation, our results show that for fixed $R/L$, $\xi/L\sim{}L^{-0.6}$. 
The prefactor of these curves increases with indenter radius, as has been previously shown in the limit of $L \gg R$~\cite{Miller:2008qw}.  
The nonlinear scaling of $\xi$ with $L$ implies that no length-scale free continuum model, suggested by the simple linear scaling of $Y_{*}$ and $D_{*}$, can possibly predict the spatial structure of the embryo.  

\section{Mesoscale analysis of incipient dislocation}
\label{section: meso analysis}

\subsection{Rationale for mesoscale analysis}
\label{subsection: mesoscale rationale}

A system about to nucleate a dislocation pair has a critical normal mode which has strong jumps (high $\Omega$) in only a small portion of the crystal. 
This motivates mesoscale analysis of the incipient defect where we `freeze out' those degrees of freedom outside some some circular \emph{meso} region and analyze its linear elastic properties. 
In particular, we compute the lowest energy eigenvalue, $\lambda_{\text{min}}$, and eigenmode of meso-regions of increasing radii centered at the embryo, in configurations close to instability ($\delta{}D\sim{}10^{-6}$).
MR showed that the lowest eigenvalue of a meso-region \emph{large enough to completely encompass the true embryo} is an excellent predictor of homogeneous dislocation nucleation~\cite{Miller:2008qw}.
Here, we systematically investigate what information can be gleaned from analysis of meso-regions \emph{smaller} than $\xi$.
 
\begin{figure}[h]
\begin{center}
\includegraphics[width=.50\textwidth]{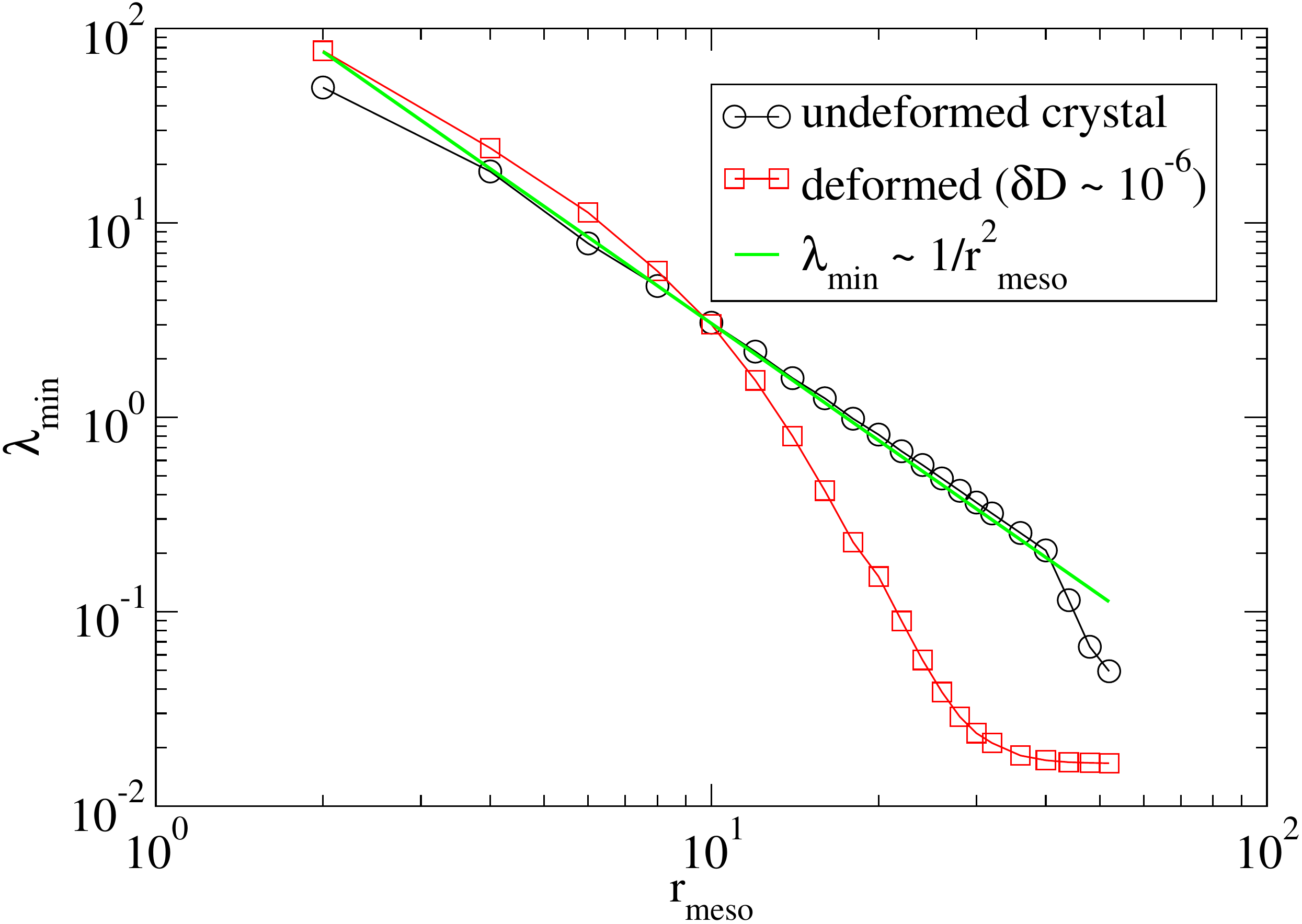} 
\includegraphics[width=.23\textwidth]{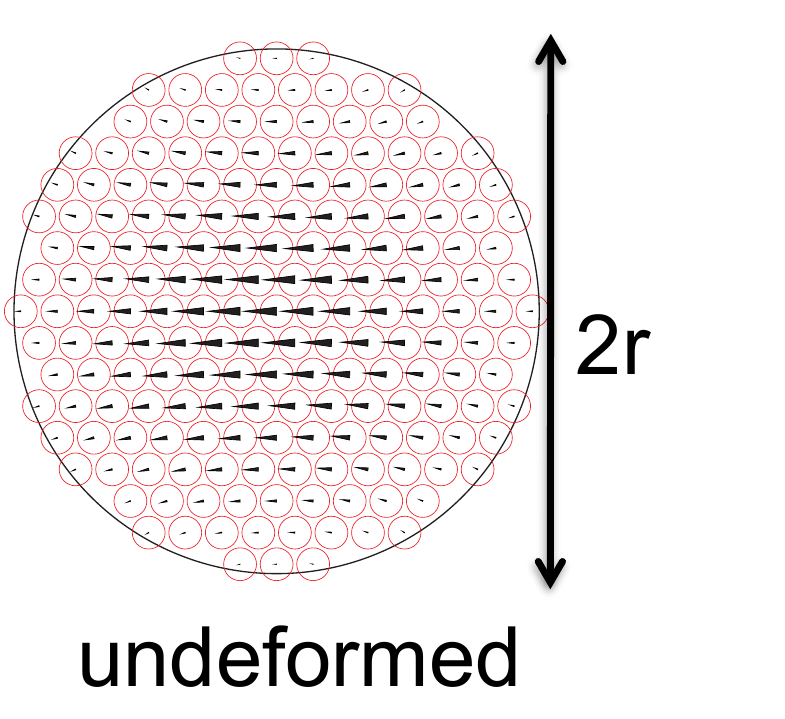}
\includegraphics[width=.22\textwidth]{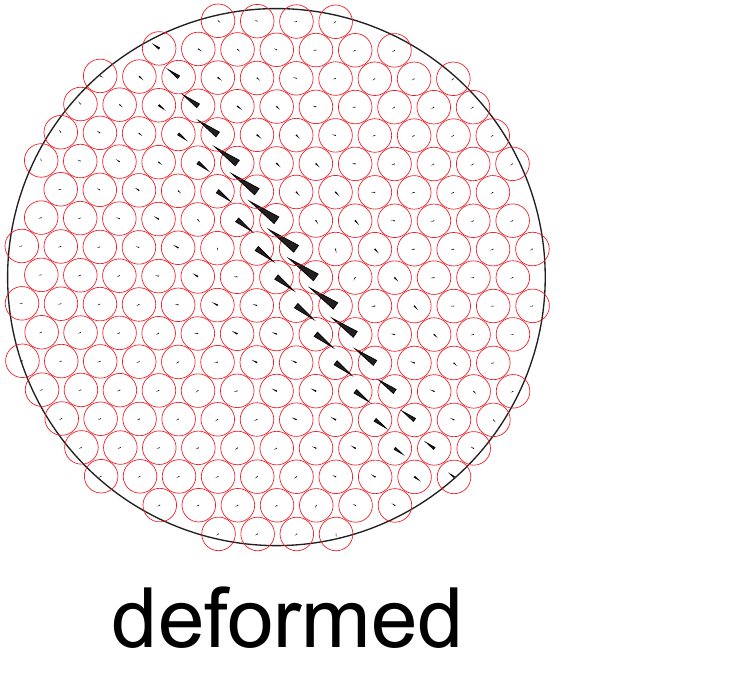} 
\end{center}
\caption{
Top:  Characteristic $\lambda_{\text{min}}$ versus $ r_{\text{meso}}$ curves for an undeformed and a deformed crystal close to nucleation ($\delta{}D\sim{}10^{-6}$).   
Bottom:  Meso region normal modes corresponding to the lowest energy eigenvalue of meso-region with radius $r_{\text{meso}} = 8$: from an undisturbed crystal (left) and a configuration close to dislocation nucleation (right).
}
\label{fig: meso analysis} 
\end{figure}

For a homogeneous, linear, elastic continuum with fixed boundaries, the energy eigenmodes scale as $1/L^{2}$ (where $L$ is a characteristic measure of system length). 
This is true for atoms in a crystal as long as the domain size is large enough that continuum effects dominate the behavior of long wavelength normal modes and the crystal atoms are confined to small displacements. 
In the top panel of figure~\ref{fig: meso analysis} we demonstrate the validity of this approximation for an undeformed crystal by showing that: $\lambda_{\text{min}}\sim{}1/r^{2}_{\text{meso}}$. 
However, the same plot shows that, for a typical deformed configuration close to nucleation ($\delta{}D\sim{}10^{-6}$),  $\lambda_{\text{min}}$ falls off faster than any power law with increasing meso-region radii. 
\footnote{The $\lambda_{\text{min}}$ versus $r_{\text{meso}}$ curve for the undeformed system drops off sharply after $r_{\text{meso}}>40$ because subsequent meso-regions hit the top surface and are no longer circular.}

\begin{figure}[t]
\begin{center}
\includegraphics[width=.5\textwidth]{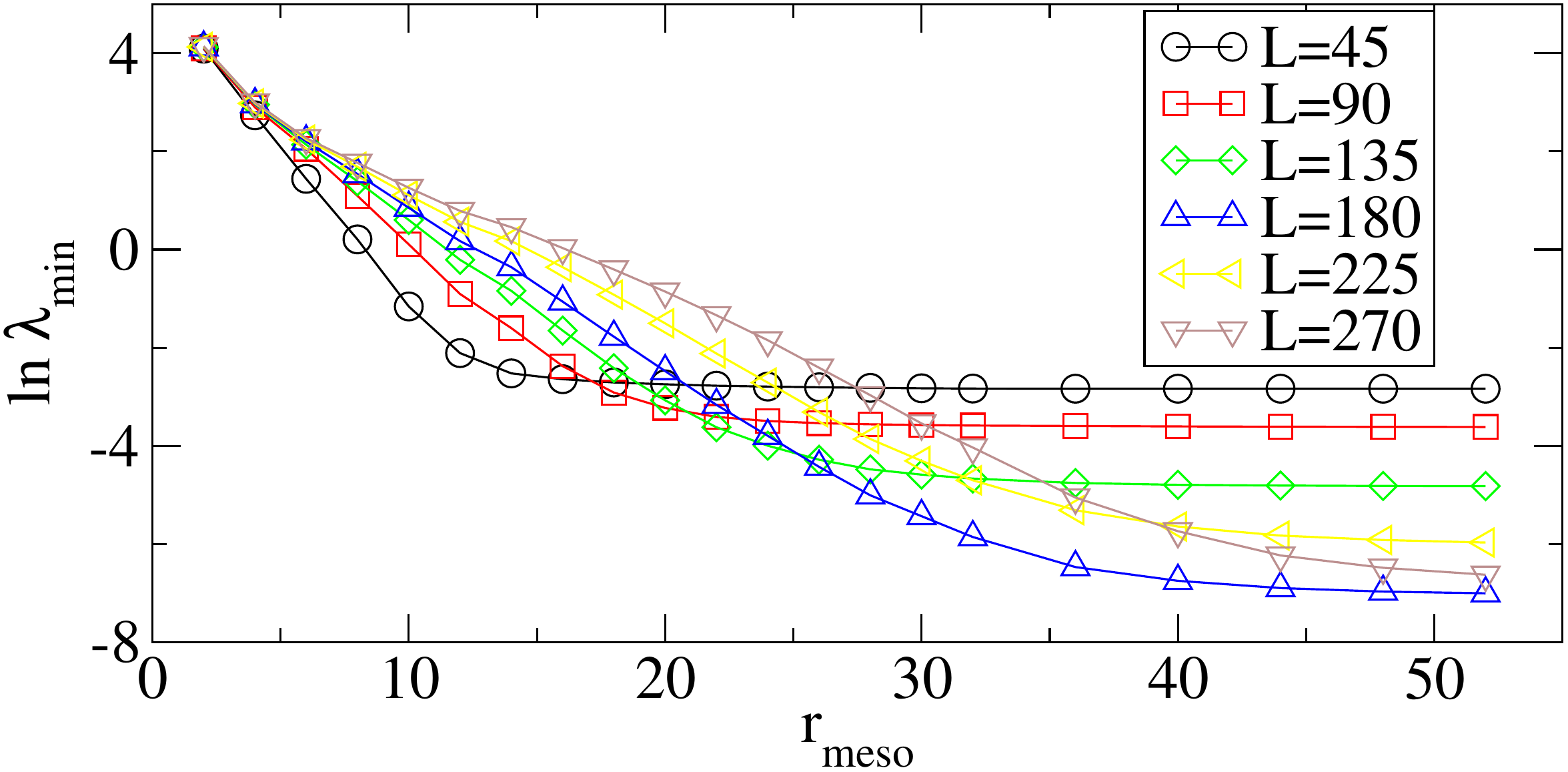} 
\end{center}
\caption{  ln$(\lambda_{\text{min}})$ versus $ r_{\text{meso}}$ curves for various film thicknesses, $L$, and one of our geometries with $R_{\text{indenter}}=2L$.
}
\label{fig: meso eig curves}
\end{figure}

For a given meso-region of radius $r_{\text{meso}}$, the normal modes corresponding to $\lambda_{\text{min}}$ for the undeformed configuration and the deformed configuration about to nucleate a dislocation are very different. 
The bottom image of figure~\ref{fig: meso analysis}, shows that the lowest energy normal mode for the \emph{undeformed} configuration, at $r_{\text{meso}}=8$, is a long wavelength mode with dominant wavelength equal to roughly twice the diameter of the meso-region (as expected from a continuum approximation). 
However for the same $r_{\text{meso}}$, the adjacent image shows that the lowest normal mode for a typical deformed configuration has a \emph{Gaussian} embryo resembling the critical normal mode shown in figure~\ref{fig: omega field} (top). 
This suggests that the \emph{structure} of the lowest meso-region normal mode has information about the true dislocation embryo, even when the eigenvalue is far from zero. 
 
Once $r_{\text{meso}}$ becomes large enough that the entire embryo is encompassed, one gains no new information by considering larger meso-regions.
Hence we expect that the $\lambda_{\text{min}}$ versus $r_{\text{meso}}$ curve for a system near nucleation must plateau at some $r_{\text{meso}}$.
The crossover to the plateau must be dependent on the embryo size $\xi$. 
In figure~\ref{fig: meso eig curves} we show ln$(\lambda_{\text{min}})$ versus $r_{\text{meso}}$ curves for one of our representative indenter-crystal geometries ($R=2L$), for configurations close to nucleation ($\delta{}D\sim{}10^{-6}$). 
We see that for deformed configurations close to nucleation, $\lambda_{min}$ rapidly decays with $r_{\text{meso}}$ before plateauing. 
In each case, there is a roughly exponential drop of $\lambda_{\text{min}}$ before it plateaus at some large $r_{\text{meso}}$.

The plateau height, the value of $\lambda_{\text{min}}$ in the limit of large $r_{\text{meso}}$, depends only on $\delta D$: as we have shown in figure~\ref{fig: lowest 4 eigenvalues}, the plateau height will go to zero as $\sqrt{\delta{}D}$. 
The crossover $r_{\text{meso}}$ increases with $L$ in accord with figure~\ref{fig: xi scaling}.
It may be tempting to think of the exponential decay of $\lambda_{\text{min}}$ with $r_{\text{meso}}$ as intrinsic property~\cite{Delph:2010zm} and to extract a length scale associated with the embryo. 
However, it should be kept in mind that the plateau value of $\lambda_{\text{min}}$  is a function of $\delta{}D$ and not an intrinsic property of the system.
Systems prepared closer to the bifurcation point, at smaller $\delta D$, have a lower plateau value and a more rapid decay to the plateau, so there is nothing intrinsic about the decay rate.

\subsection{Mesoscale analysis of critical mode structure}
\label{subsection: meso crit mode analysis}

Given that dislocation nucleation is a nonlocal phenomenon with a spatially extended embryo, the central question we address with mesoscale analysis in this paper is:  how big a region in a crystal is big enough for detecting an impending nucleation event and the associated embryo? 
Unlike earlier work \cite{Miller:2008qw,Delph:2009vf,Delph:2010zm} that focussed on the lowest energy eigenvalue, $\lambda_{\text{min}}$, of meso-regions, we analyze the the \emph{spatial structure} of the lowest normal mode of the meso-region. 
The main reason for our choice is that the height of the plateau of $\lambda_{\text{min}}$ versus $r_{\text{meso}}$ depends very sensitively on $\delta{}D$ whereas the spatial structure of the critical normal mode does not. 
Again, in all our analysis we have been careful to choose configurations where $\delta{}D$ is small enough that the lowest eigenmode of the full system corresponds to the dislocation embryo.

\begin{figure}[ht]
\begin{center}
\includegraphics[width=0.5\textwidth]{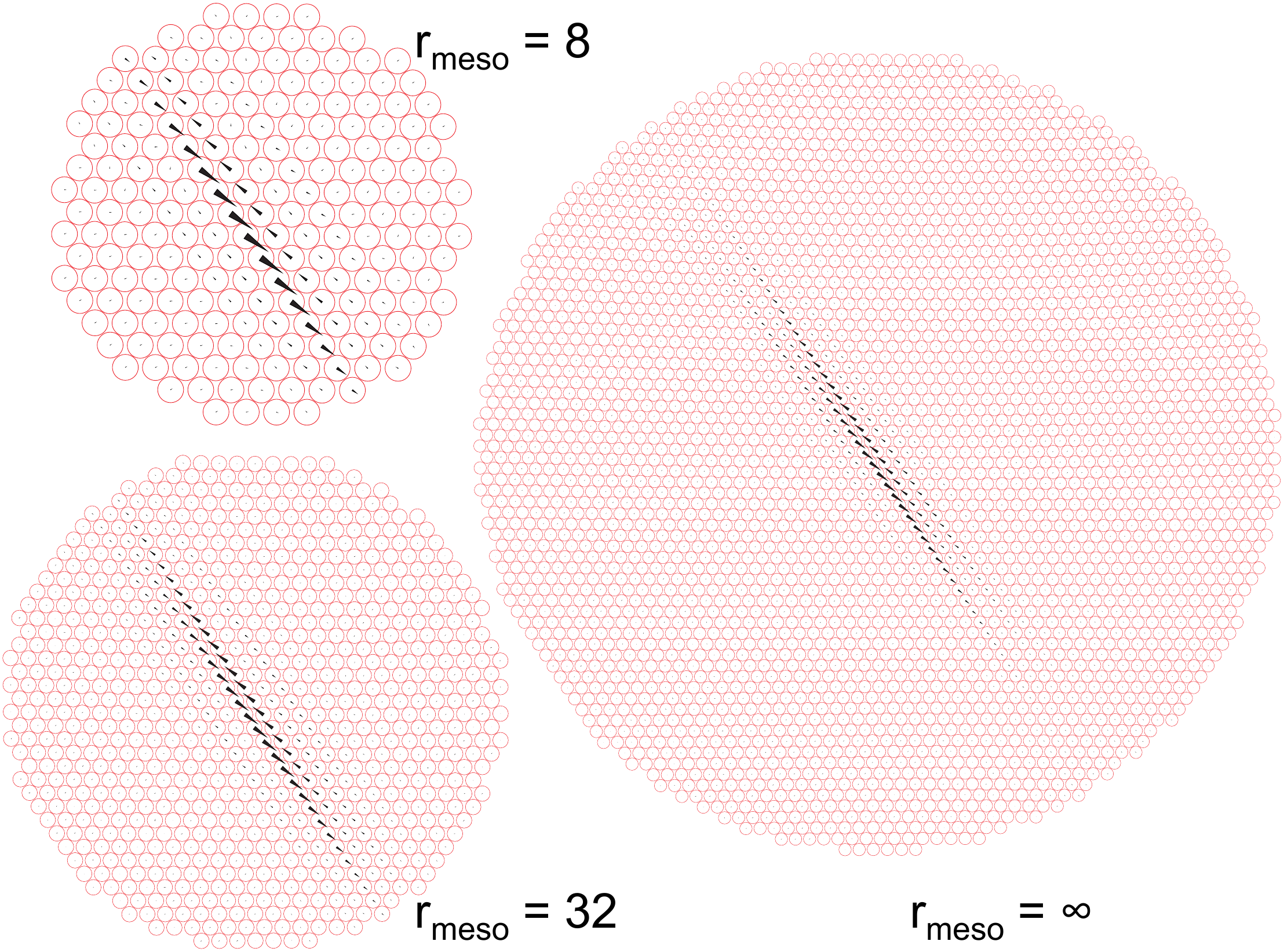}  \\
\end{center}
\caption{ 
Structure of the lowest energy eigenvalue normal mode for 3 different $r_{\text{meso}}$ from the system with $L=135$, $R=3L$ and $\delta{}D\sim{}10^{-6}$. $r_{\text{meso}}=\infty$ corresponds  to the entire system.  Note how quickly the meso normal mode captures the structure of the incipient defect.  }
\label{fig: meso mode structure}
\end{figure}

First, in figure~\ref{fig: meso mode structure}, we show the lowest normal mode in a meso-region for a typical system near its nucleation threshold. 
For comparison, the true critical mode for the full system is also shown.
We indicate the usual unconstrained eigenvalue analysis of the full system by $r_{\text{meso}}=\infty$. 
The lowest normal mode in each meso-region has a Gaussian $\Omega(s)$ profile like the true critical mode.
Note that for the system studied in figure~\ref{fig: meso mode structure}, $r_{\text{meso}}=8$  is \emph{smaller} than the spatial extent  of the slip-plane $\xi \approx 13$ measured from the $\Omega(s)$ profile of its critical mode. 
So, we conclude that even meso-regions much smaller than the intrinsic dislocation embryo capture the center and essential structure of the true critical mode.

\begin{figure}[h]
\begin{center}
\includegraphics[width=0.5\textwidth]{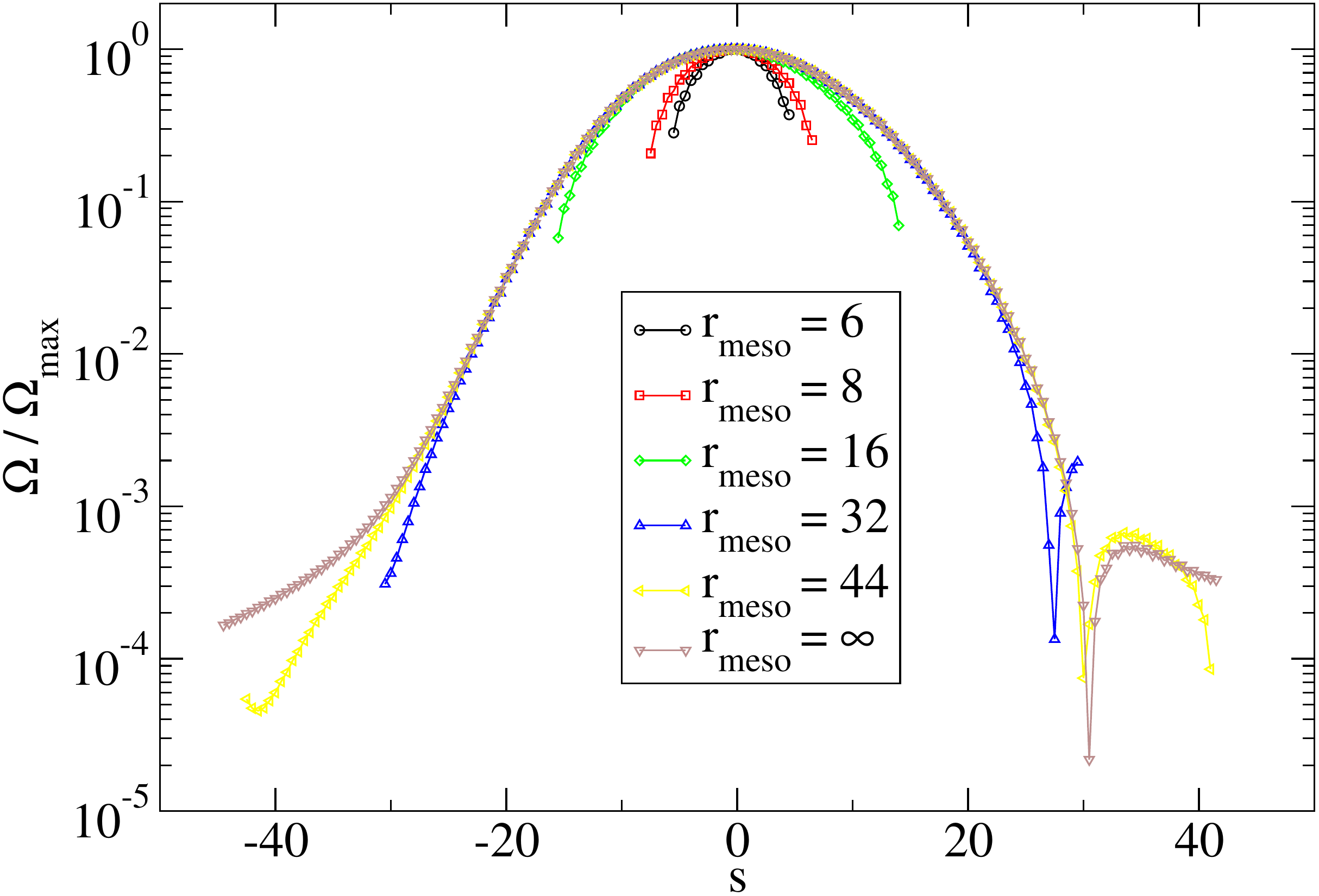} 
\end{center}
\caption{ $\Omega(s)$ curves on slip-plane computed from lowest normal modes of various mesoscale regions from the  same system as figure~\ref{fig: meso mode structure}. 
}
\label{fig: meso mode omega curves}
\end{figure}

\begin{figure}[t]
\begin{center}
\includegraphics[width=0.5\textwidth]{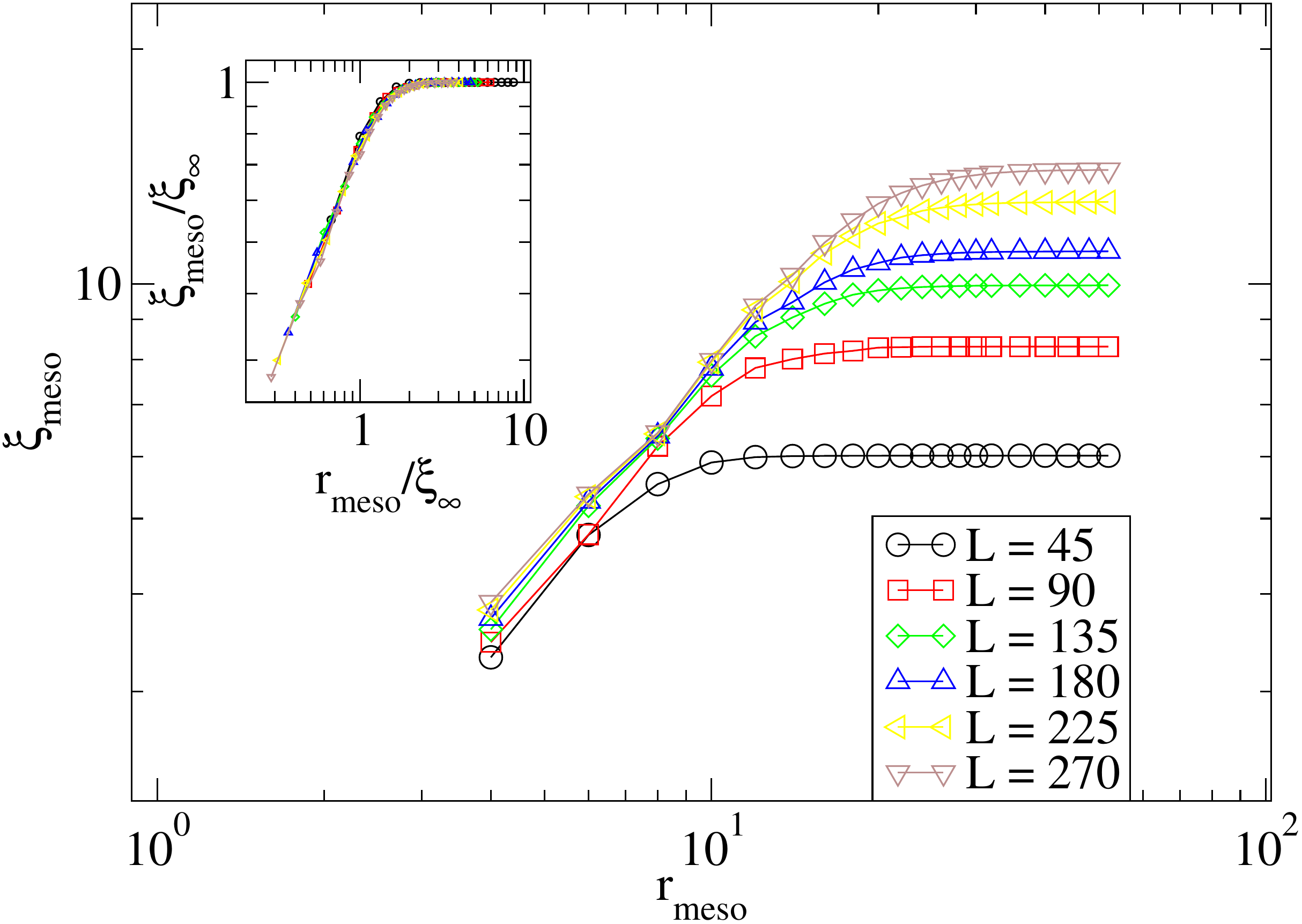}  \\
\end{center}
\caption{ 
Top: $\xi_{\text{meso}}$ as a function of $r_{\text{meso}}$ for various film thicknesses, $L$, and one of our geometries with $R_{\text{indenter}}=2L$. 
Inset shows that the curves can be made to collapse by rescaling the axes with their plateau value $\xi_{\infty}$.
}
\label{fig: meso omega curvatures against meso radii}
\end{figure}

\begin{figure}[t!]
\begin{center}
\includegraphics[width=0.5\textwidth]{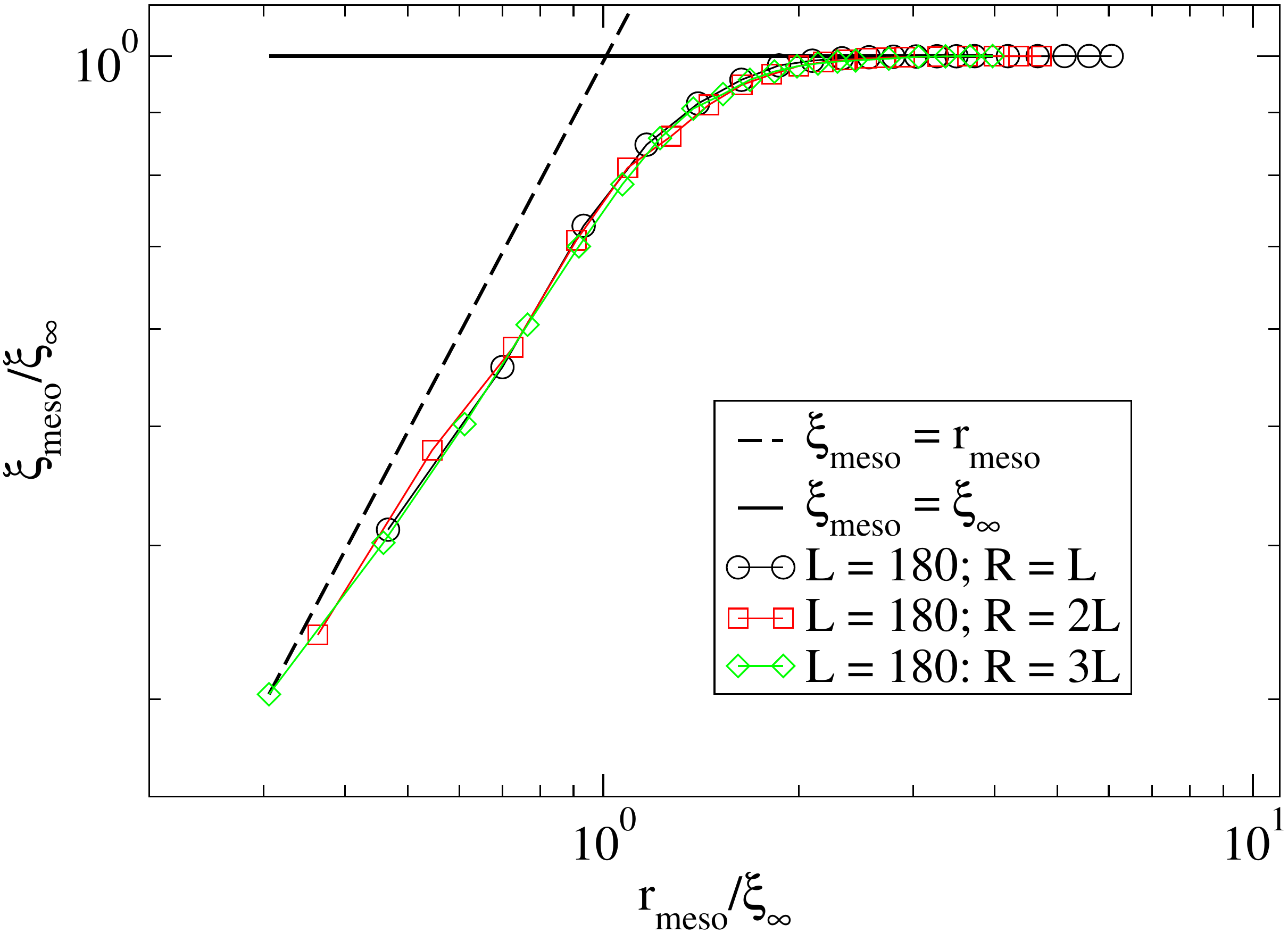}
\end{center}
\caption{ Collapsed $\xi_{\text{meso}}$ versus $r_{\text{meso}}$ curves (rescaled by $\xi_{\infty}$) for systems with $L=180$ but the 3 different indenter geometries ($R=L$, $R=2L$ and $R=3L$), showing the existence of a \emph{universal} curve. $\xi_{\text{meso}}$ initially grows sublinearly with $r_{\text{meso}}$ before plateauing at $\xi_{\infty}$
}
\label{fig: meso omega curvatures collapsed - all systems}
\end{figure}

In figure~\ref{fig: meso mode omega curves}, we show $\Omega(s)$ profiles for the lowest modes of meso-regions of increasing size.
These curves converge toward the $\Omega(s)$ profile of the true embryo as  $r_{\text{meso}}$  grows.
To quantify the spatial extent of the lowest mode in the meso-regions, in precisely the same way as for the true critical mode, we fit each $\Omega(s)$ profile to a Gaussian to extract a length scale, $\xi_{\text{meso}}$.
We label the embryo size obtained from the true critical mode in the full-system eigenmode analysis as $\xi_{\infty}$. 

Figure~\ref{fig: meso omega curvatures against meso radii} shows $\xi_{\text{meso}}$ versus $r_{\text{meso}}$ for a representative indenter-crystal geometry ($R=2L$). 
As we might expect from figure~\ref{fig: meso mode omega curves}, for small $r_{\text{meso}}$, $\xi_{\text{meso}}$ grows before plateauing at $\xi_{\infty}$. 
The increase of $\xi_{\text{meso}}$ with $r_{\text{meso}}$ for meso regions smaller than $\xi_{\infty}$ approximately follows a power law. 
This increase is expected in the small $r_{\text{meso}}$ regime where $r_{\text{meso}} < \xi_{\infty}$. 
Recall that $\xi_{\infty}$ grows with $L$ as shown in figure~\ref{fig: xi scaling}. 

In the inset of figure~\ref{fig: meso omega curvatures against meso radii} we show the rescaled curves $\xi_{\text{meso}}/\xi_{\infty}$ versus $r_{\text{meso}}/\xi_{\infty}$. 
The collapse of the rescaled curves is also excellent for the other indentation geometries, $R=L$ and $R=3L$ (not shown).  
Furthermore, in figure~\ref{fig: meso omega curvatures collapsed - all systems}, we show three rescaled curves, one from each of the three indentation geometries at $L=150$.
The $\xi_{\text{meso}}$ vs $r_{meso}$ curves, upon rescaling by $\xi_{\infty}$,  is \emph{universal} and does not depend on $R$ or $L$.
The solid black line corresponds to $\xi_{\text{meso}}/\xi_{\infty} = 1$ and the dashed black line is the $\xi_{\text{meso}}=r_{\text{meso}}$ guide line.
The increase of $\xi_{\text{meso}}$ with $r_{\text{meso}}$ is slower than linear for small meso-regions.
So, although even small meso-regions host a lowest mode that resembles a dislocation embryo, is only for  $r_{\text{meso}}\gtrsim{} 1.5\xi_\infty$, that the meso-region gives a reliable characterization of the embryo size.
It is in this sense that we say that nucleation is a quasi-local phenomenon.

\section{Discussion and Summary}
\label{section: discussion}

In this paper we studied the nucleation of dislocation dipoles in the bulk of perfect 2D crystals subjected to nanoindentation with a circular, atomistic indenter under athermal, quasistatic conditions. 
We first described the kinematics of homogeneous dislocation nucleation, showing that nucleation involves driving a system to instability along a single, critical, normal mode.
We demostrated the presence of scalings expected of such a saddle-node bifurcation type instability. 
The critical normal mode was shown to be nonlocal with strong jumps, largely confined to a pair of adjacent crystal planes. 
We introduced a measure, $\xi$, for the spatial extent of the embryo and showed that for fixed indenter radius, $R$,  $\xi$ grows with $L$, while for fixed $R/L$, $\xi\sim{}L^{0.4}$.
This universal scaling has an important consequence.
On simple dimensional grounds, no continuum theory containing only $R$ and $L$ as geometrical parameters (as used in, {\it e.g.}, reference~\cite{Li:2004JMPS} can be used, even in principle, to describe the embryonic size scaling.
 
We then performed a mesoscale analysis of configurations at the stability threshold, showing that small, nonlocal regions of the crystal, centered at the embryo core, contain significant information about an incipient nucleation event. 
However, unlike previous work that utilized the \emph{minimum eigenvalue} of the meso-regions as the main analysis tool \cite{Miller:2008qw,Delph:2009vf,Delph:2010zm}, we focused our attention on the \emph{spatial structure} of the lowest meso-region normal mode. 
We found that the relation between $\xi_{\text{meso}}$, the embryonic size inferred from the meso-region, and $r_{\text{meso}}$, the size of the meso-region itself is universal.
The lowest normal mode and eigenvalue were found to provide excellent estimates of the structure and energy of the true critical mode, but only for meso-regions larger than $r_{\text{meso}} > 1.5\; \xi$. 
This scenario leads us to think of homogeneous dislocation nucleation as quasi-local: full information about the nature of the embryo can only be obtained by analyzing sufficiently large regions, however, its existence can be inferred by examining regions much much smaller than its intrinsic size. 

The ultimate use of the analysis presented here would be to inform coarser-grained models, that do not explicitly take into account the atomic degrees of freedom, about the creation of new dislocations out of the void.
For example, in field dislocation mechanics~\cite{Acharya:2001fdm,Acharya:2010inroads} one introduces a continuous field to represent the dislocation density.
It is our hope that a criterion for dislocation nucleation based on a meso-scale analysis like we presented here could serve as a guide for the introduction of atomistic details at the dislocation embryo in concurrent multi-scale schemes built on field theories like FDM.

In the future we will check to see if the basic picture put forward here remains the same in other crystallographic orientations and in 3D with more realistic interaction potentials for various metals.
In the spirit of building a practical criterion for predicting homogeneous dislocation nucleation, an obvious extension of our mesoscale analysis is to consider meso-regions centered at different points in the crytsal and not just at the embryo center, since it is not known \emph{a priori} where a dislocation will nucleate.  
Our preliminary results (not presented here) indicate that a meso-region need only have \emph{some} overlap with the true embryo core in order to have a lowest normal mode indicate the presence of the embryo and the location of its core. 
Although our previous results~\cite{Hasan:2012bif} showed that the approach to nucleation does not appreaciably affect the structure of the critical normal mode, more work is needed to understand the effects of the proximity to nucleation, $\delta{}D$, on mesoscale analysis, and an initial study of the impact of $\delta D$ is already underway.


\acknowledgments
We thank David Rodney and Amit Acharya for useful discussions at various stages during this work.
This material is based upon work supported by the National Science Foundation under Award Number CMMI-1100245.

\bibliography{HDNRefs}
\bibliographystyle{apsrev}
\end{document}